\documentstyle[10pt,epsfig]{article}
\begin{document}
\renewcommand{\thefootnote}{\fnsymbol{footnote}}
\sloppy
\newcommand{\rp}{\right)}
\newcommand{\lp}{\left(}
\newcommand{\E}{{\rm E}}
\newcommand \be  {\begin{equation}}
\newcommand \bea {\begin{eqnarray} \nonumber }
\newcommand \ee  {\end{equation}}
\newcommand \eea {\end{eqnarray}}

\begin{center}
\centering{\bf \Large Critical Crashes}
\end{center}
\begin{center}
\centering{Anders Johansen$^1$ and Didier Sornette$^{1,2,3}$\\
\vskip 1cm
{\it $^1$ Institute of Geophysics and Planetary Physics\\
University of California, Los Angeles, California 90095\\
$^2$ Department of Earth and Space Science\\
University of California, Los Angeles, California 90095\\
$^3$ Laboratoire de
Physique de la Mati\`ere Condens\'ee\\
CNRS UMR6622 and Universit\'e des
Sciences, B.P. 70, Parc Valrose\\ 06108 Nice Cedex 2, France\\
}
}
\end{center}

\vskip 4cm

{\bf Abstract}\,:  The authors argue that the word ``critical'' in the title
is not purely literary. Based on their and other previous work on nonlinear
complex dynamical systems, they
summarize present evidence, on the Oct. 1929, Oct. 1987, Oct. 1987 Hong-Kong,
Aug. 1998 global market events
and on the 1985 Forex event, for the hypothesis advanced three years ago
that stock market crashes are caused by the slow buildup
of long-range correlations between traders leading to a collapse of the
stock market in one
critical instant.

\newpage
\pagenumbering{arabic}

\section{Crashes are outliers}

It is well-known that
the distributions of stock market returns exhibit ``fat tails'',
deviating significantly from the time-honored Gaussian description\,: a
$5\%$ dayly loss occurs approximately once every two years while the
Gaussian framework
would predict one such loss in about a thousand year.

Crashes are of an even
more extreme nature. We have measured the number $N(D)$ of times a given
level of drawn down $D$
has been observed in this century in the Dow Jones daily Average
\cite{Outlier}.
A draw down is defined as
the cumulative loss from the last local maximum to the next minimum. $N(D)$ is
well fitted by an exponential law
\be
N(D) = N_0 ~e^{-D/D_c}~,  ~~~~{\rm with} ~D_c \approx 1.8 \%~.
\label{zae}
\ee
However, we find that three events stand out blatantly. In chronological order:
World War 1 (the second largest),
Wall Street Oct. 1929 (the third largest) and Wall Street Oct. 1987 (the
largest). Each of
these draw down lasted about three days. Extrapolating the exponential fit
(\ref{zae}),
we estimate that the return time of a draw down equal to or larger than
$28.8\%$
would be more than $160$ centuries. In contrast, the market
has sustained two such events in less than a century.
This suggests a natural unambiguous
definition for a crash, as an {\it outlier}, i.e., an extraordinary event
with an amplitude above $\approx 15\%$ \cite{Outlier}.

Large price movements are often modeled as Poisson-driven jump processes. This
accounts for the bulk of the statistics. However,
the fact that large crashes are outliers implies that they are probably
controlled by different amplifying factors, which can lead to observable
precursory signatures. Here, we propose that large stock market crashes are
analogous to ``critical points'', a technical term in Physics which refers
to regimes of large-scale cooperative behavior such as close to the Curie 
temperature when the millions of tiny magnets in a bar magnet
start to influence each other and eventually end up all pointing in one
direction. We present the theory and then test it against facts.

\section{A rational imitation model of crashes}

Our model contains the following ingredients \cite{JLS}\,:
\begin{enumerate}
\item A system of traders who are influenced by their ``neighbors'';
\item Local imitation propagating spontaneously into global cooperation;
\item Global cooperation among traders causing crash;
\item Prices related to the properties of this system.
\end{enumerate}

The interplay between the progressive strengthening of imitation
controlled by the three first ingredients and the ubiquity of
noise requires a stochastic description. A crash is not certain but can
be characterized by its hazard rate $h(t)$, i.e., the probability per unit
time that the crash will happen in the next instant if it has not happened yet.

The crash hazard rate $h(t)$ embodies subtle uncertainties of the market\,:
when will the traders realize with sufficient clarity that the market is
overvalued? When will a significant fraction of them believe that the
bullish trend
is not sustainable? When will they feel that other traders think that a
crash is coming?
Nowhere is Keynes's beauty contest analogy more relevant than in the
characterization of the crash hazard rate, because the survival of the
bubble rests on the
overall confidence of investors in the market bullish trend.

A crash happens when a large group
of agents place sell orders simultaneously. This group of agents must create
enough of an imbalance in the order book for market makers to be unable to
absorb the other side without lowering prices substantially. One curious fact
is that the agents in this group typically do not know each other. They did
not convene a meeting and decide to provoke a crash. Nor do they take orders
from a leader. In fact, most of the time, these agents disagree with one
another, and submit roughly as many buy orders as sell orders (these are all
the times when a crash {\em does not} happen). The key question is to
determine by what
mechanism did they suddenly manage to organise a coordinated sell-off?

We propose the following answer \cite{JLS}\,: all the traders in the
world are organised
into a network (of family, friends, colleagues, etc) and they influence each
other {\em locally} through this network\,: for instance, an active
trader is constantly on the phone exchanging information and opinions with a
set of selected colleagues. In addition, there are indirect interactions
mediated
for instance by the media. Specifically, if I am directly
connected with $k$ other traders, then there are only two forces that
influence my opinion: (a) the opinions of these $k$ people and of the global
information network; and (b) an
idiosyncratic signal that I alone generate. Our working assumption here is that
agents tend to {\em imitate} the opinions of their connections.
The force (a) will tend to create order, while
force (b) will tend to create disorder. The main story here
is a fight between order and disorder. As far as asset prices are
concerned, a crash happens when order wins (everybody has the same
opinion: selling), and normal times are when disorder wins (buyers and
sellers disagree with each other and roughly balance each other out). We
must stress that this is exactly the opposite of the popular
characterisation of crashes as times of chaos. Disorder, or a balanced
and varied opinion spectrum, is what keeps the market liquid in normal times.
This mechanism does not require an overarching
coordination mechanism since macro-level coordination can
arise from micro-level imitation and it relies on a
realistic model of how agents form opinions by constant interactions.

In the spirit of ``mean field'' theory of collective systems
\cite{Goldenfeld},
the simplest way to describe an imitation
process is to assume that the hazard rate $h(t)$ evolves according to the
following
equation\,:
\be
{dh \over dt} = C~h^{\delta}~,~~~~~~~{\rm with}~\delta > 1~,
\label{azzer}
\ee
where $C$ is a positive constant.
Mean field theory amounts to embody the diversity of trader actions by a
single effective representative behavior determined from an average
interaction between the
traders. In this sense, $h(t)$ is the collective result of the interactions
between
traders. The term $h^{\delta}$ in the r.h.s. of (\ref{azzer}) accounts for
the fact that the
hazard rate will increase or decrease due to the presence of {\em
interactions} between the traders. The exponent $\delta > 1$ quantifies the
effective number
equal to $\delta - 1$ of interactions felt by a typical trader. The condition
$\delta > 1$ is crucial to model interactions and is, as we now show, essential
to obtain a singularity (critical point) in finite time.
Indeed, integrating (\ref{azzer}), we get
\be
h(t) = {B \over (t_c - t)^{\alpha}}~,~~~~~~~~{\rm with}~\alpha \equiv {1
\over \delta - 1}~.
\label{cjdjlk}
\ee
The critical time $t_c$ is determined by the initial conditions at some
origin of time.
The exponent $\alpha$ must lie between zero and one for an economic
reason\,: otherwise, as we shall see,
the price would go to infinity when approaching $t_c$ (if the bubble has
not crashed in the mean time). This condition translates into $2 < \delta <
+\infty$\,: a typical trader must be
connected to more than one other trader. There is a large body of literature in
Physics, Biology and Mathematics on the microscopic modeling of systems of
stochastic dynamical
interacting agents that lead to critical behaviors of the type
(\ref{cjdjlk}) \cite{Liggett}. The macroscopic
model (\ref{azzer}) can thus be substantiated by specific microscopic
models \cite{JLS}.

The critical time $t_c$ signals the death of the speculative bubble.
We stress that $t_c$ is not {\em the} time of the crash because the crash
could happen at any time before $t_c$, even though this is not very likely.
$t_c$ is the most probable time of the crash. There
exists a finite probability
\be
1- \int_{t_0}^{t_c} h(t) dt >0
\ee
of  ``landing'' smoothly, i.e. of
 attaining the end of the bubble without crash. This residual probability is
crucial for the coherence of the model, because otherwise agents would
anticipate the crash
and not remain in the market.

Assume for simplicity that, during a crash, the price drops
by a fixed percentage $\kappa\in(0,1)$, say between $20$ and $30\%$ of the
price
increase above a reference value $p_1$.
Then, the dynamics of the asset price before the crash are given by:
\be
\label{eq:crash}
dp = \mu(t)\,p(t)\,dt-\kappa [p(t)-p_1] dj~,
\ee
where $j$ denotes a jump process whose value is zero before the crash and one
afterwards. In this simplified model,
we neglect interest rate, risk aversion, information asymmetry,
and the market-clearing condition.

As a first-order approximation of the market organization, we assume that
traders
do their best and price the asset so that a fair game condition holds.
Mathematically, this stylized rational expectation model
is equivalent to the familiar martingale hypothesis:
\be
\label{eq:martingale}
\forall t'>t \qquad\  E_t[p(t')] = p(t)
\ee
where $p(t)$ denotes the price of the asset at time $t$ and $\E_t[\cdot]$
denotes the expectation conditional on information revealed up to time $t$.
If we do not allow the asset price to fluctuate under the impact of noise,
the solution to Equation (\ref{eq:martingale}) is a constant: $p(t) = p(t_0)$,
where $t_0$ denotes some initial time. $p(t)$ can be interpreted as the
price in
excess of the fundamental value of the asset.

Putting (\ref{eq:crash}) in (\ref{eq:martingale}) leads to
\be
\mu(t) p(t) = \kappa [p(t)-p_1] h(t)~.
\label{hfjqklq}
\ee
In words, if
the crash hazard rate $h(t)$ increases, the return $\mu$ increases to
compensate the traders
for the increasing risk. Plugging (\ref{hfjqklq}) into
(\ref{eq:crash}), we obtain a ordinary differential equation. For
$p(t) - p(t_0) < p(t_0) - p_1$, its solution is
\be
\label{eq:price}
p(t) \approx p(t_0) +  \kappa [p(t_0) - p_1]~\int_{t_0}^t h(t') dt'
\qquad\mbox{before the crash}.
\ee
This regime applies to the relatively short time scales
of two to three years prior to the crash shown below.

The higher the probability of a crash, the faster the price must increase
(conditional on having no crash) in order to satisfy the martingale (no
free lunch) condition.
Intuitively, investors must be compensated by the chance of a higher return
in order to be induced to hold an asset that might crash. This effect
may go against the naive preconception that price is adversely affected by
the probability of the crash, but our result is the only one consistent with
rational expectations.

Using (\ref{cjdjlk}) into (\ref{eq:price}) gives the
following price law:
\be
\label{eq:solution}
p(t) \approx p_c -\frac{\kappa B}{\beta}\times(t_c-t)^{\beta}
\qquad\mbox{before the crash}.
\ee
where $\beta = 1-\alpha\in(0,1)$ and $p_c$ is the price at the critical
 time (conditioned on no crash having been triggered). The price before the
crash
follows a power law with a finite upper bound $p_c$.
The trend of the price becomes unbounded as we
approach the critical date. This is to compensate for an unbounded
crash rate in the next instant.

\section{Log-periodicity}

The last ingredient of the model is to recognize that
the stock market is made of actors which differs in
size by many orders of magnitudes ranging from individuals to gigantic
professional investors, such as pension funds. Furthermore, structures at
even higher levels, such as currency influence spheres (US\$, Euro, YEN ...),
exist and with the current globalisation and de-regulation of the market
one may argue that structures on the largest possible scale, i.e.,
the world economy, are beginning to form. This means that the structure
of the financial markets have features which resembles that of hierarchical
systems with ``traders'' on all levels of the market. Of course, this
does not imply that any strict hierarchical structure of the stock market
exists, but there are numerous examples of qualitatively hierarchical
structures in society. Models \cite{JLS,SJ98} of imitative interactions on
hierarchical
structures recover the power law behavior (\ref{eq:solution}). But in
addition,
they predict that the critical exponent $\alpha$ can be a
complex number!
The first order expansion of the general solution for the hazard rate is then
\be
\label{eq:hazard3}
h(t)\approx B_0(t_c-t)^{-\alpha}
+B_1(t_c-t)^{-\alpha}\cos[\omega\log(t_c-t)+\psi ].
\ee
Once again, the crash hazard rate explodes near the critical date. In
addition,
it now displays log-periodic oscillations. The evolution of the
price  before the crash and before the critical date is given by:
\be
\label{eq:complex}
p(t) \approx p_c -\frac{\kappa}{\beta}\left\{
B_0(t_c-t)^{\beta}
+B_1(t_c-t)^{\beta}\cos[\omega\log(t_c-t)+\phi]\right\}
\ee
where $\phi$ is another phase constant. The key feature is that oscillations
appear in the price of the asset before the critical date. The local
maxima of the function are separated by time intervals that tend to zero at the
critical date, and do so in geometric progression, i.e., the ratio of
consecutive time
intervals is a constant
\be
\lambda \equiv e^{2 \pi \over \omega}~.
\ee
This is very useful from an empirical point
of view because such oscillations are much more strikingly visible in actual
data than a simple power law\,: a fit can ``lock in'' on the oscillations which
contain information about the critical date $t_c$.
Note that complex exponents and log-periodic oscillations do
not necessitate a pre-existing hierarchical structure as mentioned above, but
may emerge spontaneously from the non-linear
complex dynamics of markets \cite{Revue}.

In Natural Sciences, critical points are widely considered to be one of the
most interesting properties of complex systems. A system goes critical when
local influences propagate over long distances and the average state of the
system becomes
exquisitely sensitive to a small perturbation, i.e., different parts of
the system becomes highly correlated. Another characteristic is that
critical systems are self-similar across scales: in our example, at the
critical point, an ocean of traders who are mostly bullish may have within
it several
islands of traders who are mostly bearish, each of which in turns surrounds
lakes of bullish traders with islets of bearish traders; the progression
continues all the way down to the smallest possible scale: a single trader
\cite{Wilson}. Intuitively speaking, critical self-similarity is why
local imitation cascades through the scales into global coordination.

\section{Fitting the crashes}

Details on our numerical procedure are given in \cite{JLS}. Figures 1-3
show the behavior of the market index prior to the four
crashes of Oct. 1929 (Fig.1), Aug. 1998,  Oct. 1997 (Hong-Kong) (Fig.2) and
of Oct. 1987 (Fig.3). In addition, Fig. 3 shows the US \$ expressed in
DEM and CHF currencies before the collapse of the bubble in 1985. A fit
with Eq.~(\ref{eq:complex}) is shown as a continuous line for each event.
The table summarises the key parameters. Note the small fluctuations in 
the value of the scaling ratio $2.2 \leq \lambda \leq 2.7$ for the 4 
stock market crashes. This agreement constitutes one of the key test of our 
theory. Rather remarkably, the scaling ratio for the DEM and CHF currencies 
against the US\$ is comparable.

\begin{table*}[h]
\begin{center}
\begin{tabular}{|c|c|c|c|c|c|c|c|c|c|} \hline
crash & $t_c$ & $t_{max}$ & $t_{min}$ & $\%$ drop & $\beta$ & $\omega$ &
$\lambda$ \\ \hline
1929 &  $30.22$ & $29.65$ & $29.87$ & $46.9\%$ & $0.45$ & $7.9$ & $2.2$\\
\hline
1985 (DEM) &  $85.20$ & $85.15$ & $85.30$ & $14\%$ & $0.28$ & $6.0$ & $2.8$
\\ \hline
1985 (CHF) &  $85.19$ & $85.18$ & $85.30$ & $15\%$ & $0.36$ & $5.2$ & $3.4$
\\ \hline
1987 &  $87.74$ & $87.65$ & $87.80$ & $29.7\%$ & $0.33$ & $7.4$ & $2.3$\\
\hline
1997 (H-K) &  $97.74$ & $97.60$ & $97.82$ & $46.2\%$ & $0.34$ & $7.5$ &
$2.3$\\ \hline
1998 &  $98.72$ & $98.55$ & $98.67$ & $19.4\%$ & $0.60$ & $6.4$ & $2.7$\\
\hline
\end{tabular}
\end{center}
Table\,: $t_c$ is the critical time predicted from the fit of the market
index to
the Eq.~(\ref{eq:complex}). The other parameters $\beta$, $\omega$ and
$\lambda$
of the fit are also shown. The fit is performed up to the time $t_{max}$
at which the market index achieved its highest maximum before the crash.
$t_{min}$
is the time of the lowest point of the market. The percentage drop is
calculated from the total loss from $t_{max}$ to $t_{min}$.

\end{table*}

In order to investigate the significance of these results,
we picked at random fifty  $400$-week intervals in the period
1910 to 1996 of the Dow Jones average and launched the fitting procedure
described in \cite{JLS} on these surrogate data sets.
The results were very encouraging. Of
the eleven fits with a quality of fit comparable with that of the other
crashes, only six data sets produced values for $\beta$ and $\omega$
which were in the same range. All six fits belonged to the periods prior to the
crashes of 1929, 1962 and 1987. The existence of a ``crash'' in 1962 was
before these results unknown to us and the identification of this crash
naturally strengthens the case. We refer the reader to \cite{JLS} for a
presentation of the best fit obtained for this ``crash''.

In the last few weeks before a crash,
the market indices shown in Fig.~1-3 depart from the final acceleration
predicted by Eq.~5\,: this is the regime where the hazard rate becomes
extremely high, the market becomes more and more sensitive to ``shocks''
and the market
idiosyncrasies are
bound to have an increasing impact. Within the theory of critical phenomena,
it is well-known that the singular behavior of the observable, here the
hazard rate or the rate of change of the stock market index, will be smoothed
out by the finiteness of the market. Technically, this is referred to as a
``finite-size effect''.

In order to qualify further the significance of the log-periodic oscillations
in a non-parametric way,
we have eliminated the leading trend from the price data by the following
transformation
\be \label{residue}
p\lp t\rp \rightarrow \frac{p\lp t\rp - \left[ p_c -
\frac{\kappa}{\beta} B_0(t_c-t)^{\beta}\right]}{\frac{\kappa}{\beta}
B_1(t_c-t)^{\beta}},
\ee
which should leave us with a pure $\cos[\omega\log(t_c-t)+\phi]$ if no other
effects were present. In figure 4, we see this residue prior to the
1987 crash with a very convincing periodic trend as a function of
$\log\lp \frac{t_c - t}{t_c}\rp$. We estimated the significance of this trend
by using a so-called Lomb periodogram for the four index crashes and the two
bubble collapse on the Forex considered here.
The Lomb periodogram is a local fit of a cosine (with a phase) using
some user chosen range of frequencies. In figure 5, we see a
peak around $f \approx 1.1$ for all six cases corresponding to
$\omega = 2\pi f \approx 7$ in perfect agreement with the previous results.
We note that only the relative level of the peak {\it for each separate}
periodogram should be regarded a measure of the significance of the
oscillations. Since the nature of the ``noise'' is unknown and very likely
different for each crash, we cannot estimate the confidence interval of
the peak and compare the results for the different crashes. We also note,
the the strength of the oscillations is $\approx 5\%$ of the leading power
law behaviour for all 6 cases signfiying that they cannot be neglegted.

\section{Towards a prediction of the next crash?}

How long time prior to a crash can one identify the log-periodic signatures?
Not only one would like to predict future crashes, but it is important to
further test how
robust our results are. Obviously, if the log-periodic structure of the
data is purely accidental, then the parameter values obtained should depend
heavily on the size of the time interval used in the fitting. We have thus
carried out a systematic testing procedure \cite{JLS} using a second order
expansion of the hazard rate \cite{SJ97} and a time interval of 8 years prior
to the two crashes of 1929 and 1987.
The general picture we obtain is the following. For the Oct. 1987 crash,
a year or more before a crash, the data is not sufficient to give any
conclusive results. Approximately a year before the crash, the fit begins
to lock-in
on the date of the crash with increasing precision and our procedure becomes
robust. However, if one
wants to actually predict the time of the crash, a major obstacle is the
fact that several possible dates are possible. In addition,
 the fit in general ``over-shoot'' the true day of the crash.
For the Oct. 1929 crash, we have to wait until
approximately $4$ month before the crash for the fit to lock in on the
date of the crash, but from that point the picture is the same as for the
crash in Oct. 1987. We caution the reader that jumping in the prediction game
may be hazardous and misleading\,: one deals with a delicate optimization
problem
that requires extensive back and forward testing. Furthermore, the formulas
given here are only ``first-order'' approximations and novel improved
methods are needed
\cite{New}. Finally, one must never forget that the crash has to remain in
part a random event
in order to exist!

A general trend for the analysis of the five crashes presented here is that
the critical $t_c$ obtained from fitting the data
tends to over-shoot the time of the crash.
This observation is fully consistent with our rational expectation model of
a crash. Indeed, $t_c$ is {\it not} the time of the crash but
the most probable value of the skewed distribution of the possible times of
the crash. The occurrence of the crash is a random phenomenon which occurs
with a
probability that increases as time approaches $t_c$. Thus, we expect that
fits will give
values of $t_c$ which are in general close to but {\it systematically}
later than the real time of the crash. The phenomenon of ``overshot'' that
we have clearly
documented \cite{JLS} is thus fully consistent with the theory.

It is a striking observation that essentially similar
 crashes have punctuated this century, notwithstanding
tremendous changes in all imaginable ways of life and work. The only thing
that has probably little changed are the way humans think and behave.
The concept that emerges here is that the organization of traders in
financial markets
leads intrinsically to ``systemic instabilities'', that probably result in
a very robust way
from the fundamental nature of human beings, including our gregarious
behavior, our
greediness, our reptilian psychology during panics and crowd behavior and
our risk aversion.
The global behavior of the market, with its log-periodic structures that emerge
as a result of the cooperative behavior of traders, is reminiscent of the
process of the emergence of intelligent behavior at a macroscopic scale
that individuals at the  microscopic scale have not idea of. This process has
been discussed in biology for instance in animal populations such as ant
colonies or in connection with the emergence of consciousness
\cite{Anderson}.

\vskip 0.5cm
\noindent anders@moho.ess.ucla.edu\\
sornette@cyclop.ess.ucla.edu

\pagebreak

\newpage

\begin{figure}
\begin{center}
\epsfig{file=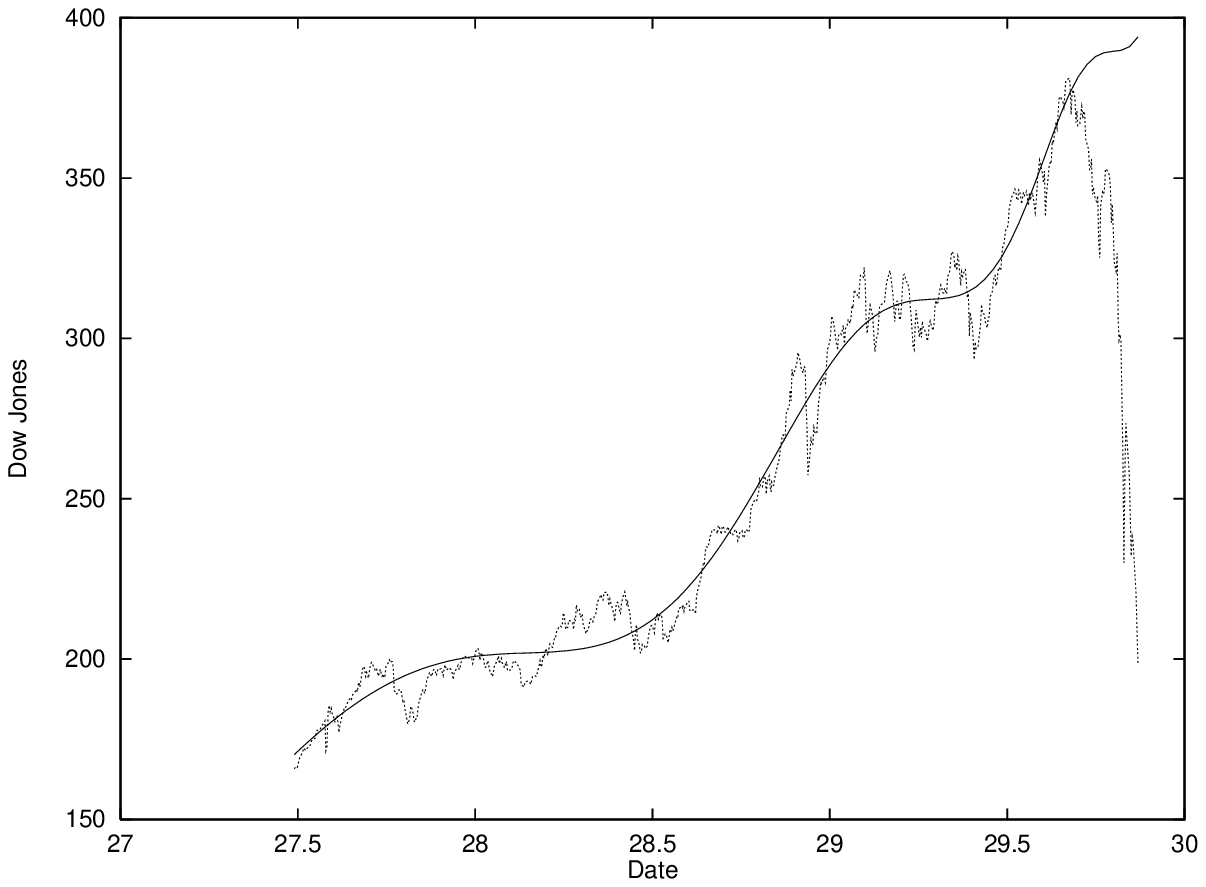,  width=15cm, height=10cm}
\caption{\protect\label{29} The Dow Jones index prior to the October 1929
crash on Wall Street. The fit is equation (\protect\ref{eq:complex}) with
$p_c \approx  571 $, $ \frac{\kappa}{\beta}B_0\approx -267 $, $
\frac{\kappa}{\beta}B_1\approx 14.3 $, $\beta\approx  0.45 $, $
t_c \approx  30.22$, $ \phi\approx  1.0 $, $ \omega\approx  7.9$.}
\end{center}
\end{figure}

\begin{figure}
\begin{center}
\epsfig{file=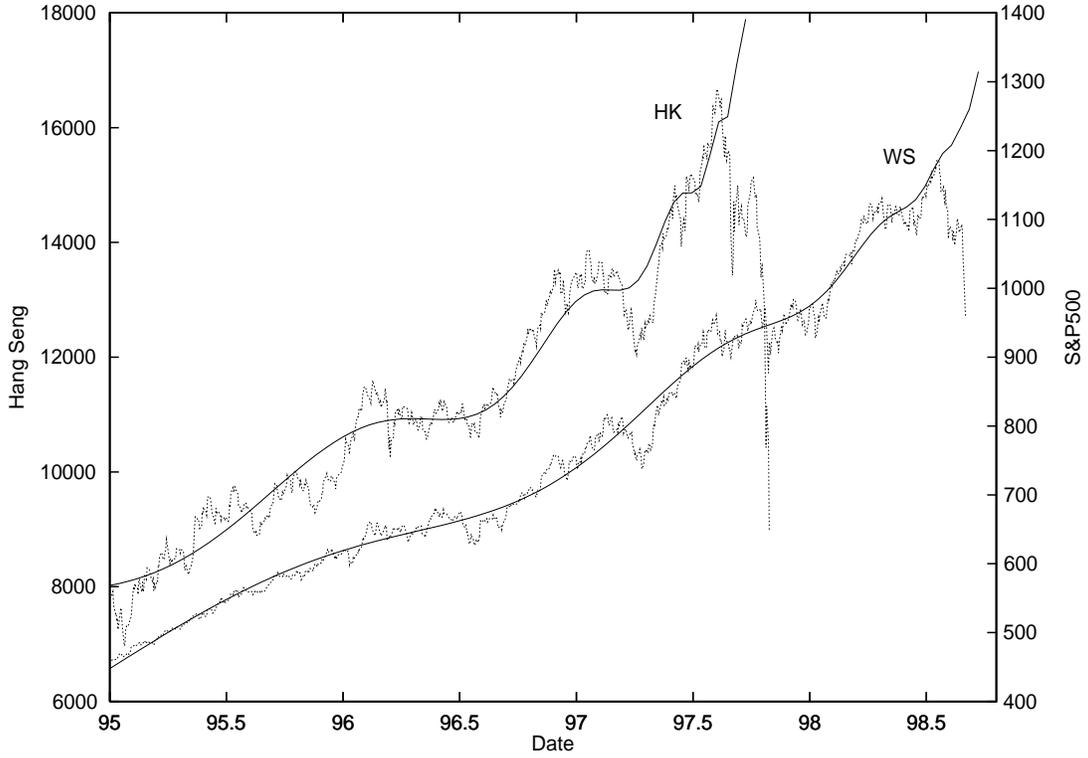,  width=15cm, height=10cm}
\caption{\protect\label{98} The Hang Seng index prior to the October 1997
crash on the Hong-Kong Stock Exchange and the S\&P 500 stock market index
prior to the
recent crash on Wall Street in August 1998. The fit to the Hang Seng index
is equation (\protect\ref{eq:complex})
with $p_c \approx  20077 $, $ \frac{\kappa}{\beta}B_0\approx  -8241 $, $
\frac{\kappa}{\beta} B_1\approx  -397$, $\beta\approx  0.34 $, $ t_c\approx
97.74 $, $ \phi\approx  0.78 $, $ \omega\approx  7.5$.
The fit to the S\&P 500 index is equation (\protect\ref{eq:complex})
with $p_c \approx  1321$, $ \frac{\kappa}{\beta}B_0\approx -402 $, $
\frac{\kappa}{\beta}B_1 \approx  19.7 $, $ \beta \approx  0.60$, $
 t_c\approx  98.72 $, $ \phi \approx  0.75$, $ \omega \approx  6.4$.}
\end{center}
\end{figure}

\begin{figure}
\begin{center}
\epsfig{file=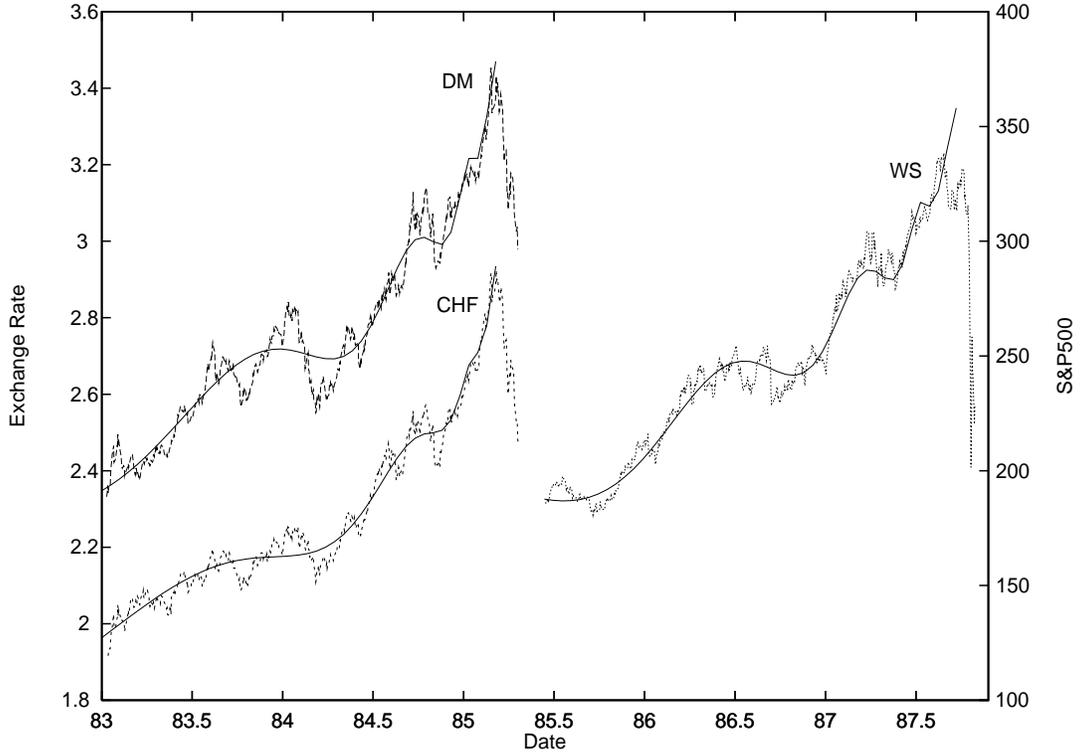,  width=15cm, height=10cm}
\caption{\protect\label{87} The S\& P 500 US index prior to the October
1987 crash on Wall Street and the US \$ against DEM and CHF prior to the
collapse mid-85. The fit to the S\&P 500 is equation (\protect\ref{eq:complex}) with
$p_c\approx  412  $, $ \frac{\kappa}{\beta}B_0\approx  -165 $, $
\frac{\kappa}{\beta}B_1 \approx   12.2 $, $  \beta \approx   0.33 $, $
t_c \approx   87.74  $, $ \phi \approx   2.0 $, $ \omega \approx   7.4$.
The fits to the DM and CHF currencies against the US dollar gives
$p_c\approx  3.88  $, $ \frac{\kappa}{\beta}B_0\approx  -1.2 $, $
\frac{\kappa}{\beta}B_1 \approx   0.08 $, $  \beta \approx   0.28 $, $
t_c \approx   85.20  $, $ \phi \approx   -1.2 $, $ \omega \approx   6.0$ and
$p_c\approx  3.1  $, $ \frac{\kappa}{\beta}B_0\approx  -0.86 $, $
\frac{\kappa}{\beta}B_1 \approx   0.05 $, $  \beta \approx   0.36 $, $
t_c \approx   85.19  $, $ \phi \approx   -0.59 $, $ \omega \approx   5.2$,
respectively.
}
\end{center}
\end{figure}

\begin{figure}
\begin{center}
\epsfig{file=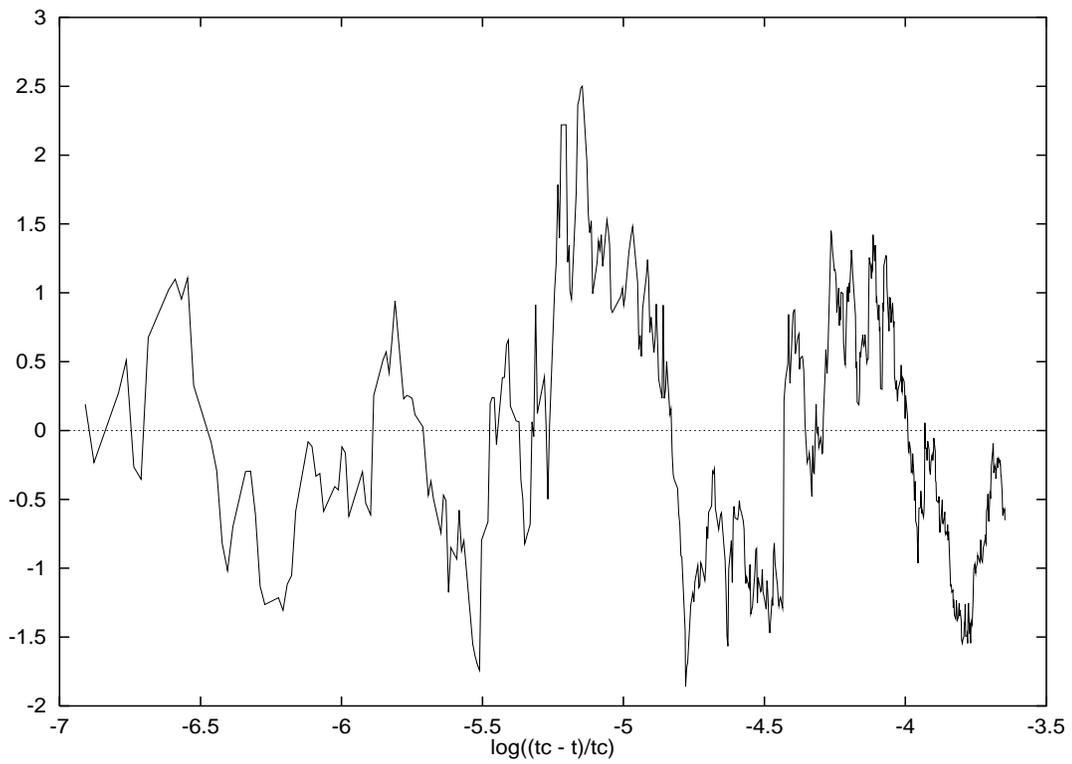,  width=15cm, height=10cm}
\caption{\protect\label{lp87} The residue as defined by the transformation
(\protect\ref{residue}) as a function of $\log\lp \frac{t_c - t}{t_c}\rp$ for
the 1987 crash.}
\end{center}
\end{figure}

\begin{figure}
\begin{center}
\epsfig{file=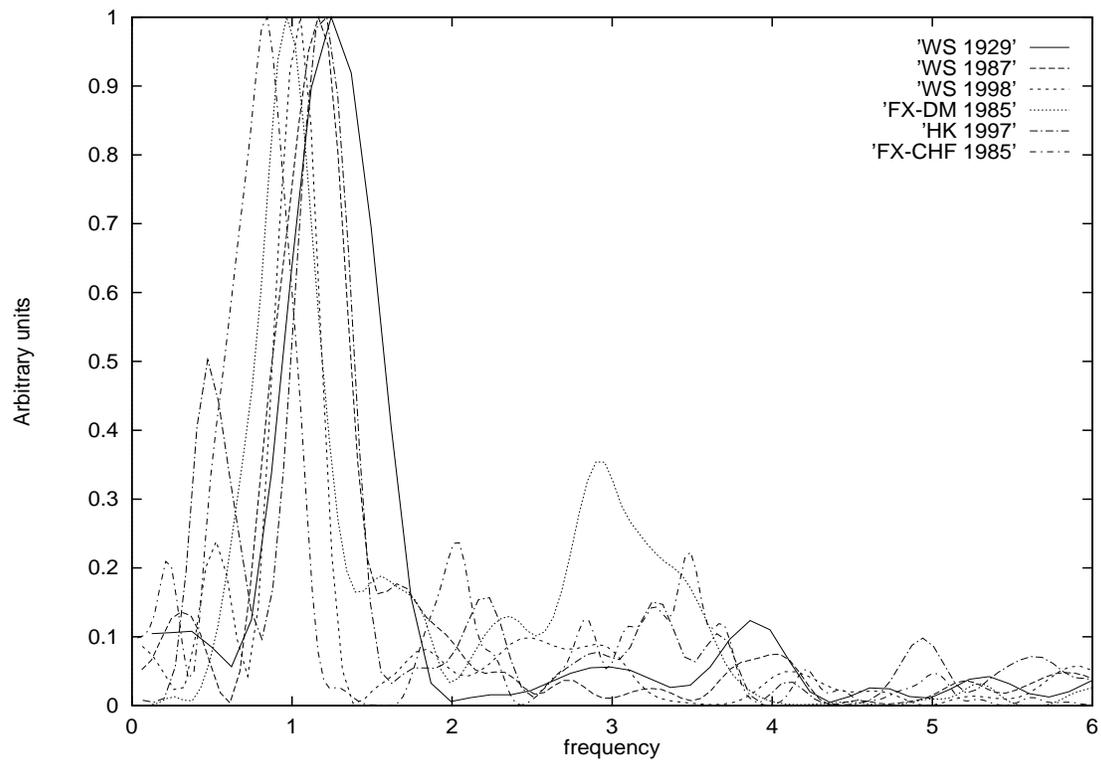, width=15cm, height=10cm}
\caption{\protect\label{lombcrash} The Lomb periodogram for the 1929, 1987
and 1998 crashes and Wall Street, the 1997 crash on the Hong Kong
Stock Exchange and the 1985 US \$ currency crash in 1985.
For each periodogram, the significance of the peak should be
estimated against the noise level.}
\end{center}
\end{figure}

\end{document}